\documentclass[review]{elsarticle}

\usepackage{lineno,hyperref}
\usepackage{tikz-cd}
\usepackage{dsfont}
\usepackage{amsfonts}
\usepackage{amsmath}
\usepackage{slashed}
\usepackage{graphics}
\usepackage{amsmath}
\usepackage{amssymb}
\usepackage{tikz-cd}
\usepackage[compat=1.1.0]{tikz-feynman}
\usepackage{setspace}
\usepackage{float}
\usepackage{cleveref}
\usepackage{units}
\modulolinenumbers[5]

\journal{Universe (MDPI)}
\begin{document}

\begin{frontmatter}

\title{\textbf{Neutrino oscillations and Lorentz Invariance Violation}}

\author[Unimi]{M.D.C. Torri\corref{mycorrespondingauthor}}
\cortext[mycorrespondingauthor]{Corresponding author}
\ead{marco.torri@unimi.it, marco.torri@mi.infn.it}

\address[Unimi]{Dipartimento di Fisica, Universit\'a degli Studi di Milano and INFN Milano\\via Celoria 16, I - 20133 Milano, Italy}

\begin{abstract}
This work explores the possibility to resort to neutrino phenomenology to detect evidences of new physics, caused by the residual signals of the supposed quantum structure of space-time. In particular this work will investigate the effects on neutrino oscillations and mass hierarchy detection, predicted by models that violates Lorentz invariance, preserving the space-time isotropy and homogeneity. Neutrino physics is the ideal environment where conducting the search for new "exotic" physics, since the oscillation phenomenon is not yet completely included in the Standard Model of particles and this can require the necessity to investigate new theoretical models.  Moreover LIV theories are constructed assuming a modified kinematics, caused by the interaction of massive particles with the space-time background. This means that the dispersion relations are modified, so it appears natural to search for effects caused by this physics in phenomena governed by masses, such in the case of neutrino oscillations. Finally in neutrino oscillations three different mass eigenstates are involved and in a LIV scenario that preserves isotropy at least two different species of particle must interact.\\
\end{abstract}


\end{frontmatter}

%%%%%%%%%%%%%%%%%%%
\setcounter{section}{0} %% Remove this when starting to work on the template.

\section{Introduction}
Recent observations made by experiments with natural (solar) neutrino sources \cite{solar1,solar2,solar3,solar4,solar5,solar6,solar7}, atmospheric \cite{atmospheric}, artificial neutrinos short baseline \cite{short-reactors1,short-reactors2,short-reactors3,short-reactors4,short-reactors5,short-reactors6} and long baseline reactor neutrinos \cite{KamLAND1,KamLAND2,LBL-acceleratori1,LBL-acceleratori2,LBL-acceleratori3} confirm the existence of the neutrino flavor oscillation phenomenon. The oscillation evidence have been further reinforced by the discussed LSND \cite{LSND1,LSND2} and MiniBOONE \cite{MiniBoone1,MiniBoone2} experiments and even by the appearance experiments, like the CNGS beam \cite{OPERA}, T2K \cite{T2K} and No$\nu$a \cite{NOVA}, which collect neutrino signals with changed flavor respect to the produced beam. It is well known that this new physics can not be explained by the minimal particle physics Standard Model (SM), where only $3$ left handed massless neutrino flavors are included. This new physics effect is usually described by supposing the existence of tiny neutrino masses, that can cause the oscillations. This produces a model ($3\nu$SM extension of the Standard Model of particle physics, that includes the $3$ neutrino masses) where the oscillations are governed by a $3\times3$ matrix, determined by $6$ parameters, $3$ angles $\theta_{12},\,\theta_{23}$ and $\theta_{13}$, a phase $\delta$ that takes into account CP violation in weak interaction and $2$ mass squared differences, which depend on the neutrino mass hierarchy. Neutrinos appear therefore the ideal candidates to search for new "exotic" physical effects. In this work only the search for new physics caused by Lorentz Invariance Violation (LIV) is considered in both oscillation and mass hierarchy detection.

\section{LIV models}
In the past many attempts to extend the SM of particle physics have been conducted. Some extensions introduce an additional symmetry between bosons and fermions, i.e. supersymmetry. Other models look for an extension of the standard gauge group $SU(3)\times SU(2)\times U(1)$ into a more general symmetry group, which reduces to the classical one via a spontaneous breaking mechanism that produces the standard physics scenario. All these theories are based on Lorentz Invariance (LI). This symmetry is nowadays at the root of our understanding of nature. Even if there is no definitive evidence to sustain departures from LI, there are consistent points indicating that Lorentz Invariance Violation (LIV) can be consequence of quantum gravity. There are therefore consistent motivations to conduct systematic tests of this fundamental symmetry validity. Neutrino physics seems to be the ideal environment where to conduct this physical research, since $3$ different mass eigenstates are involved in the oscillation process. In fact, to detect possible LIV effects in an isotropic scenario it is necessary that at least $2$ different particles species interact.\\
The prevalent means used to search for LIV effects consists in formulating Effective Field Theories (EFT) extensions of the SM of particle physics, in order to obtain phenomenological predictions that can be experimentally tested. The principal EFT beyond the SM are Very Special Relativity (VSR) and the Standard Model Extension (SME). These models share the common feature of being based on highly reasonable assumptions deemed appropriate to test LI in every possible sector. \\
The first EFT approach to LIV considered was introduced by Coleman and Glashow \cite{Glash1,Glash2}. They developed an isotropic perturbative framework to deal with LIV departures from classical quantum field theories, modifying the Lagrangian such in a way that causes the maximum attainable velocities of massive particles to differ from the speed of light $c$. The perturbations are conceived so that the gauge symmetry $SU(3)\times SU(2)\times U(1)$ is preserved. Moreover this kind of perturbations are rotationally and translationally invariants, but in a preferred fixed inertial frame.\\
The most complete and coherent EFT framework to study the LIV phenomenology is referred as Standard Model Extension (SME) \cite{Koste1,Koste2}. This theory explores the LIV scenarios by amending the particle SM, supplementing all the possible LIV operators, that preserve the gauge symmetry $SU(3)\times SU(2)\times U(1)$. The SME formulation is conceived even in order to preserve microcausality, positive energy and four-momentum conservation law. Moreover the quantization principles are conserved, in order to guarantee the existence of Dirac and Schr\"odinger equations, in the correct energy regime limit.
Therefore the SME modifications consist of perturbation operators, generated by the coupling of matter Lagrangian standard fields with background tensors. These tensors non-zero void expectation value and their constant non dynamical nature break the LI under active transformations of the observed system. It is important to underline that this model introduce a difference between active and passive reference frame transformations, not present in Special Relativity (SR) \cite{Kosteproc,Kosteproc2,Kosteproc3}. Active transformations refer to the transformation that affect the observed particle, instead the passive ones are those that affect the observer. The presence of couplings with a fixed background induces a Lorentz violation only for active transformations, that modify the coupling of the observed particle field with the background tensors. In this sense SME preserves the covariance of phisics formulation under passive transformations, that is observer rotations or boosts.\\
Another approach to LIV consists in attempting the construction of complete physical theories, such as Doubly Special Relativity (DSR) \cite{Amelino,Amelino2,Amelino3,Amelino4,Smolin1,Smolin2}. The main motivation for constructing such a new theory consists in the attempt to reconcile the existence of a second universal constant (the Planck lenght) with the relativity principle, because the distance contraction induced by the Lorentz transformations is in contrast with the idea of a minimum invariant length. This approach to LIV in its last formulation is known as \emph{Relative Locality} \cite{AmelinoCamelia}. The central idea of this model consists in supposing the momentum space and not the space-time as the fundamental structure to describe physics. Space-time is considered only a local projection of the momentum space. The new proposal of this model is that the concept of absolute locality is relaxed and different observers feel a personal space-time structure, which is energy (or equivalently momentum) dependent. This model is based on simple semiclassical assumptions about the momentum space geometry, that determine departure from the classical space-time description, first of all the relativity principle is modified and acquires a local character.\\
The new \emph{Principle of local relativity} states that the momentum space is the fundamental structure at the basis of the physical processes description, instead space-time description is constructed by every observer in a personal, local way, loosing universality. Space-time becomes therefore an auxiliary concept, which emerges from the fundamental momentum space, where the real dynamics takes place\cite{Gutierrez,Pfeiffer,Ballesteros}.

\section{HMSR - Homogeneously Modified Special Relativity}
This model is constructed in the attempt of preserving isotropy and homogeneity of space-time in a LIV scenario \cite{TorriHMSR,TorriUHECR}. To pursue this aim the interaction with the background is geometrized. In fact in this theory Dispersion Relations (DR) are modified to perturb the kinematic, in order to geometrize the interaction of massive particles with the supposed quantum structure of the background.
\begin{equation}
\label{c1}
MDR(p):=\,E^2-\left(1-f\left(\frac{|\overrightarrow{p}|}{E}\right)-g\left(\frac{\overrightarrow{p}}{E}\right)\right)|\overrightarrow{p}|^2=m^2
\end{equation}
the $f$ function is constructed to preserve the MDR rotational invariance. Moreover the Modified Dispersion Relation (MDR) does not present a dependence on particle helicity or spin, in fact it is constructed without distinctions between particles and antiparticles, so the constructed theory is CPT even. Since the publication of the Greenberg theorem \cite{Greenberg}, it is recognized that LIV does not imply CPT violation. In the same work the opposite statement was declared true, but this point is widely debated in literaure \cite{Dolgov-et-al,Discussion-LIV-CPT,Discussion-LIV-CPT2,Discussion-LIV-CPT3,Discussion-LIV-CPT4}.\\
Promoting the MDR to the role of norm in the momentum space, it is possible to obtain the momentum space Finsler metric:
\begin{equation}
\label{c14}
\widetilde{g}^{\mu\nu}(p)=\left(
                             \begin{array}{cc}
                                1 & 0 \\
                                0 & -(1-f(p/E))\mathbb{I}_{3\times3} \\
                             \end{array}
                          \right)
\end{equation}
The associated metric of the coordinate space can be obtained via the Legendre transformation:
\begin{equation}
\label{c22}
g(x,\,\dot{x}(p))_{\mu\nu}=\left(
                     \begin{array}{cc}
                         1 & 0 \\
                         0 & -(1+f(p/E))\mathbb{I}_{3\times3} \\
                     \end{array}
                  \right)\\
\end{equation}
and the associated generalized vierbein:
\begin{equation}
\label{c27}
\begin{split}
& e^{\mu}_{\,a}(p)=        \left(
                             \begin{array}{cc}
                               1 & \overrightarrow{0} \\
                               \overrightarrow{0}^{t} & \sqrt{1-f(p)}\,\mathbb{I}_{3\times3} \\
                             \end{array}
                           \right)\\ \\
& e_{\mu}^{\,a}(p)=        \left(
                             \begin{array}{cc}
                               1 & \overrightarrow{0} \\
                               \overrightarrow{0}^{t} & \sqrt{1+f(p)}\,\mathbb{I}_{3\times3} \\
                             \end{array}
                           \right)
\end{split}
\end{equation}
The resulting metric is an asymptotically flat finslerian structure \cite{Koste11,Liberati1,Edwards,Lammerzahl,Bubuianu,Schreck}. All the physical quantities are therefore generalized, acquiring an explicit dependence on the momenta. In this model every particle species has its own metric, with a personal maximum attainable velocity. For this aspect this model is a generalization of VSR, i.e. it admits VSR as an high energy limit. Moreover every particle lives in a modified curved personal space-time, therefore it is necessary to introduce a new mathematical formalism to conduct computations between physical quantities related to different interacting particles. The elements of the vierbein can be used as projectors from the local curved space to a common support Minkowski flat space-time. The graph of the transition from one tangent (local) space to the other becomes:
\[\begin{tikzcd}
         (TM,\,\eta_{ab},\,p) \arrow{d}{e(p)} \arrow{rr}[swap]{\Lambda} && (TM,\,\eta_{ab},\,p')\arrow{d}[swap]{\overline{e}(p')} \\
        (T_{x}M,\,g_{\mu\nu}(p)) \arrow{rr}[swap]{\overline{e}\circ\Lambda\circ e^{-1}} && (T_{x}M,\,\overline{g}_{\mu\nu}(p'))
\end{tikzcd}\]
where is indicated the explicit dependence of the metric from momenta. An original feature of this model consists in the possibility to construct the modified Lorentz group. Now, using again the vierbein to project physical quantities from the local to global space, it is possible to define the general Modified Lorentz Transformations (MLT) as:
\begin{equation}
\label{c40}
\Lambda_{\mu}^{\;\nu}(p)=e_{\;\mu}^{a}\,(\Lambda p)\Lambda_{a}^{\;b}\,e_{\;\nu}^{b}(p)
\end{equation}
These Modified Lorentz Transformations (MLT) are the isometries of the MDR (\ref{c1}), that is every species presents its personal MLT, which are the isometries for the MDR of the particle. The new physics, caused by LIV, emerges only in the interaction of two different species. That is every particle type physics is modified in a different way by LIV.  Therefore, to analyze the interaction of two particles, it is necessary to determine how the reaction invariants - that is the Mandelstam relativistic invariants - are modified. For this reason it is natural to generalize the definition of the internal product of the sum of two particle species momenta.
\begin{equation}
\label{c45}
\langle p+q|p+q \rangle=(p_{\mu}\,e_{a}^{\,\mu}(p)+q_{\mu}\,\tilde{e}_{a}^{\,\mu}(q))\,\eta^{ab}\,(p_{\nu}\,e_{b}^{\,\nu}(p)+q_{\nu}\,\tilde{e}_{b}^{\,\nu}(q))
\end{equation}
With this internal product it is now possible to generalize the definition of the Mandelstam variables $s$, $t$ and $u$ in such a way that can preserve the theory covariant formulation respect to the MLT. This means that it is not necessary to introduce a preferred reference frame, in contrast with the great part of the other LIV models.\\
Now it is possible to generalize the SM of particle physics \cite{TorriHMSR}, following a procedure analogous to that of SME. First of all it is necessary to modify the Dirac matrices and the related Clifford Algebra, introducing the explicit dependence on the momenta:
\begin{equation}
\begin{split}
\label{c51}
&\{\Gamma_{\mu},\Gamma_{\nu}\}=2\,g^{\mu\nu}(p)=2\,e_{\mu}^{\,a}(p)\,\eta_{ab}\,e_{\nu}^{\,b}(p)\\
&\Gamma^{\mu}=e^{\;\mu}_{a}(p)\,\gamma^{a}\qquad \Gamma_5=\frac{\epsilon^{\mu\nu\alpha\beta}}{4!}\Gamma_{\mu}\Gamma_{\nu}\Gamma_{\alpha}\Gamma_{\beta}=\gamma_5
\end{split}
\end{equation}
Next it is possible to modify the spinor fields, obtaining:
\begin{equation}
\label{c56}
\begin{split}
&\psi^{+}(x)=u_{r}(p)e^{-ip_{\mu}x^{\mu}}\\
&\psi^{-}(x)=v_{r}(p)e^{ip_{\mu}x^{\mu}}
\end{split}
\end{equation}
\begin{equation}
\label{c59}
u_{r}(m,\,\overrightarrow{0})=\chi_{r}=\left(
                                         \begin{array}{c}
                                           1 \\
                                           0 \\
                                         \end{array}
                                       \right)
\end{equation}
It is simple to demonstrate that the modified Dirac equation
\begin{equation}
\label{cc59}
(i\Gamma^{\mu}\partial_{\mu}-m)\psi=0
\end{equation}
implies the MDR (\ref{c1}).\\
Finally it is possible to obtain an amended formulation of the SM of particles, where for every field an associated vierbein is used to project the physical quantities from the modified personal curved space-time to the common Minlowski space.
As an example it is reported the explicit form of the QED Lagrangian:
\begin{equation}
\label{c68}
\mathcal{L}=\sqrt{|\det{[g]}|}\;\;\overline{\psi}(i\Gamma^{\mu}\partial_{\mu}-m)\psi+e\sqrt{|\det{[\widetilde{g}]}|}\;\;\overline{\psi}\,\Gamma_{\mu}(p,\,p')\,\psi\,\overline{e}^{\mu}_{\;\nu}\,A^{\nu}
\end{equation}
where $\overline{e}$ represents the \emph{vierbein} correlated to the gauge field and the index $\mu$ represents a coordinate of the Minkowski space-time $(TM,\,\eta_{\mu\nu})$. The term that multiplies the conserved current is a generalization of the analogous term borrowed from curved space-time QFT, where its explicit form is given by: $\sqrt{|\det{[g]}|}$. In the low energy scenario the perturbations is negligible, on the contrary in the high energy limit, it is possible to consider incoming and outgoing momenta with approximately the same magnitude, even after interaction. Therefore the conserved currents do not depend on the momenta and admit a constant form high energy limit. The definition of the conserved current reduces, as in \cite{TorriHMSR}, to:
\begin{equation}
\label{c69}
J_{\mu}=e\sqrt{|\det{1/2\{\Gamma_{\mu},\,\Gamma_{\nu}\}}|}\;\;\overline{\psi}\,\Gamma_{\mu}\,\psi=e\sqrt{|\det{[g]}|}\;\;\overline{\psi}\,\Gamma_{\mu}\,\psi
\end{equation}\\
The modified SM formulation preserves the classical gauge $SU(3)\times SU(2)\times U(1)$. In fact it is possible to demonstrate that the Coleman Mandula theorem is still valid, even if the symmetry group is given by $\mathcal{P}(p)\otimes G_{int}$, where $\mathcal{P}(p)$ is the direct product of modified Poincar\'e groups, that depends explicitly on the particle species and energy (momentum):
\begin{equation}
\label{c105a}
\mathcal{P}(p)=\otimes_{i}\mathcal{P}^{(i)}(p_{(i)})
\end{equation}
and $G_{int}$ is the internal symmetries group (in this case $SU(3)\times SU(2)\times U(1)$).\\
Even the Poincar\'e brackets are modified, in fact it is possible to obtain:
\begin{equation}
\label{h1}
\begin{split}
&\{\widetilde{x}^{\mu},\,\widetilde{x}^{\nu}\}=\{x^{i}e_{i}^{\,\mu}(p),\,x^{j}e_{j}^{\,\nu}(p)\}=\{x^{i},\,e_{j}^{\,\nu}(p)\}e_{i}^{\,\mu}(p)x^{j}+\{e_{i}^{\,\mu}(p),\,x^{j}\}x^{i}e_{j}^{\,\nu}(p)\\
&\{\widetilde{x}^{\mu},\,\widetilde{p}_{\nu}\}=\{x^{i}e_{i}^{\,\mu}(p),\,p_{j}e^{j}_{\,\nu}(p)\}=\{x^{i},\,e^{j}_{\,\nu}(p)\}e_{i}^{\,\mu}(p)p_{j}+\{e_{i}^{\,\mu}(p),\,p_{j}\}x^{i}e^{j}_{\,\nu}(p)
\end{split}
\end{equation}
where the coordinates $\widetilde{x}$ and $\widetilde{p}$ are defined in the curved space-time, modified by every particle. The coordinates $x$ and $p$ are defined on the flat local model. It is necessary to consider that the vierbein (\ref{c27}) is function of the ratio of momentum and energy, therefore the Poincar\'e brackets acquire a non trivial form as in curved momentum space theories \cite{Kowalski-Glikman}.

\section{LIV and neutrino oscillations - Hamiltonian approach}
Now it is possible to focus on the analysis of the eventual Lorentz violation effects impact on neutrino phenomenology. The introduction of LIV can in fact modify the flavor oscillation probabilities.
Following the SME approach to LIV, the extended Standard Model Lagrangian can be written in the general form  \cite{Koste2,Koste4,Koste5,Kostel3}:
\begin{equation}
\label{e1}
\mathcal{L} = \mathcal{L}_{0} + \mathcal{L}_{LIV}
\end{equation}
with
\begin{equation}
\label{e2}
\mathcal{L}_{LIV}=-(a_{L})_{\mu}\overline{\psi}_{L}\gamma^{\mu}\overline{\psi}_{L}-(c_{L})_{\mu\nu}\overline{\psi}_{L}\gamma^{\mu}\partial^{\nu}\overline{\psi}_{L}
\end{equation}
The first term, proportional to $(a_{L})$, in eq.(\ref{e2}), violates CPT and consequently the Lorentz invariance, while the second contribution, proportional to $(c_{L})$, breaks ``only" Lorentz Invariance. In this way one can write the effective Hamiltonian with the explicit form:
\begin{equation}
\label{e3}
H_{eff} = H_{0} + H_{LIV}
\end{equation}
where $H_{0}$ denotes the standard Lorentz covariant Hamiltonian and $H_{LIV}$ indicates the perturbation introduced by the LIV violating terms (\ref{e2}). Neglecting the standard part of the Hamiltonian $(H_{0})$ since it contributes identically to all the three mass eigenvalues oscillations probabilities for a fixed momentum neutrino beam, it is possible to use a perturbative approach. The remaning part of the extended Hamiltonian becomes therefore:
\begin{equation}
\label{e4}
H = \frac{1}{2E}\left(M^2+2(a_{L})_{\mu}p^{\mu}+2(c_{L})_{\mu\nu}p^{\mu}p^{\nu}\right)
\end{equation}
where $M^2$ is a $3\times3$ matrix, that in the mass eigenvalues basis assumes the form:
\begin{equation}
\label{e5}
\left(\begin{array}{ccc}
         m_{1}^{2} & 0 & 0 \\
         0 & m_{2}^{2} & 0 \\
         0 & 0 & m_{3}^{2} \\
      \end{array}\right)
\end{equation}
\\
Using the quantum mechanic perturbation theory, the new eigenstates become:
\begin{equation}
\label{e6}
|\widetilde{\nu}_{i}\rangle=|\nu_{i}\rangle+\sum_{i\neq j}\frac{\langle\nu_{j}|H_{LIV}|\nu_{i}\rangle}{E_{i}-E_{j}}|\nu_{j}\rangle
\end{equation}
Now one can introduce the perturbed time evolution operator:
\begin{equation}
\label{e7}
\begin{split}
&S(t)=\left(e^{-(iH_{0}+H_{LIV})t} e^{iH_{0}t}\right)e^{-iH_{0}t}=\\
=&\left(e^{-i(H_{0}+H_{LIV})t} e^{iH_{0}t}\right)S^{0}(t)
\end{split}
\end{equation}
and the oscillation probability can be evaluated as:
\begin{equation}
\begin{split}
\label{e8}
&P(\nu_{\alpha}\rightarrow\nu_{\beta})=|\langle\beta(t)|\alpha(0)\rangle|^2=\\
&\Biggr|\sum_{n}\left[\langle\beta(t)|\left(|n_{0}\rangle\langle n_{0}|+\sum_{j\neq n}\frac{\langle j_{0}|H_{LIV}|n_{0}\rangle}{E^{0}_{n}-E^{0}_{j}}|j_{0}\rangle\langle j_{0}|\right)|\alpha(0)\rangle+\ldots\right]\Biggr|^2\\
&=P^{0}(\nu_{\alpha}\rightarrow\nu_{\beta})+P^{1}(\nu_{\alpha}\rightarrow\nu_{\beta})+\ldots
\end{split}
\end{equation}
In eq.(\ref{e8}) $P^{0}(\nu_{\alpha}\rightarrow\nu_{\beta})$ represents the standard predicted oscillation probability, the remaining term is given by:
\begin{equation}
\label{e9}
\begin{split}
&P^{1}(\nu_{\alpha}\rightarrow\nu_{\beta})=\\
=&\sum_{ij}\sum_{\rho\sigma}
2L\,\mathfrak{Re}\left(\left(S^{0}_{\alpha\beta}\right)^{*} U_{\alpha i}U_{\rho i}^{*}H^{LIV}_{\rho\sigma}U_{\sigma j}U_{\beta j}^{*}\tau_{ij}\right)
\end{split}
\end{equation}
with:
\begin{equation}
\label{e10}
U_{\alpha i}=\langle\alpha|i\rangle
\end{equation}
where $|\alpha\rangle$ represents a generic flavor eigenstate and $|j\rangle$ denotes a mass eigenstate. Moreover in (\ref{e9}):
\begin{equation}
\begin{split}
\label{e11}
\tau_{ij}=\left\{
            \begin{array}{ll}
              (-i)e^{-iE_{i}t}\qquad i=j\\
              \frac{e^{-iE_{i}t}-e^{-iE_{j}t}}{E_{i}-E_{j}}\qquad i\neq j
            \end{array}
          \right.
\end{split}
\end{equation}
with the constrains on the Hamiltonian matrix:
\begin{equation}
\label{e12}
\left\{
  \begin{array}{ll}
    H^{LIV}_{\alpha\beta}=\left(H^{LIV}_{\beta\alpha}\right)^{*}\qquad\alpha\neq\beta \\
    H^{LIV}_{\alpha\alpha}\in\mathbb{R}
  \end{array}
\right.
\end{equation}
Hence also the flavor transition probability can be expanded perturbatively, as expected \cite{Diaz1,Diaz2,Diaz3}. Assuming a direction depending perturbation, in a general treatment of $H_{LIV}$, it would be necessary to specify a privileged frame of reference to report this kind of results. But the HMSR LIV model preserves isotropy and a privileged class of inertial observers is not required.\\
In the SME scenario is set the attempt to derive a model that can justify the neutrino oscillation phenomenon resorting to LIV: the puma model \cite{Diaz4}.
Following the Hamiltonian approach, this function is perturbed in order to take into account the perturbation induced by LIV effects:
\begin{equation}
\label{z1}
H_{LIV}=A(E)\left(
              \begin{array}{ccc}
                1 & 1 & 1 \\
                1 & 1 & 1 \\
                1 & 1 & 1 \\
              \end{array}
            \right)+
        B(E))\left(
              \begin{array}{ccc}
                1 & 1 & 1 \\
                1 & 0 & 0 \\
                1 & 0 & 0 \\
              \end{array}
            \right)+
        C(E))\left(
              \begin{array}{ccc}
                1 & 0 & 0 \\
                0 & 0 & 0 \\
                0 & 0 & 0 \\
              \end{array}
            \right)
\end{equation}
where $A(E),\,B(E)$ and $C(E)$ are functions of the energy of the particle. This model is rotationally invariant, but it is not covariant under the action of boosts. Moreover this model does not require all the parameters needed to describe neutrino oscillations used in standard description. However this theory necessitates to resort to the idea of neutrino mass to describe oscillations. For example in classical description the $3\nu$SM survival probabilities are described resorting to $4$ parameters: $\Delta m_{sol}^2,\,\theta_{12},\,\Delta m_{atm}^2,\,\theta_{13}$, instead the puma model requires only one mass parameter $m$ to describe the phenomenon.

\section{HMSR and neutrino oscillations}
Using directly the MDR constitutes an equivalent way to introduce LIV even in neutrino oscillations phenomenology \cite{Antonelli}. This is the way followed geometrizing the neutrino interactions with the background and in this work neutrino MDRs are supposed spherically symmetric. In this way the MDRs are assumed with explicit form (\ref{c1}).
Since the perturbation function $f$ is supposed homogeneous of degree $0$, the MDR is originated by a metric in the momentum space as already shown and this guarantees the validity of Hamiltonian dynamics. The ultra-relativistic particle propagation in vacuum is governed by the Schr\"odinger equation, whose solutions are written in the form of generic plane waves:
\begin{equation}
\label{e14}
e^{i(p_{\mu}x^{\mu})}=e^{i(Et-\overrightarrow{p}\cdot\overrightarrow{x})}=e^{i\phi}
\end{equation}
The effects of the modified metric do not appear, because the correction terms simplify, since the contraction is between a covariant and a controvariant vector. To give the explicit form of the solution, it is possible to start from the MDR (\ref{c1}), and using the approximation of ultrarelativistic particle $|\overrightarrow{p}|\simeq E$, we obtain:
\begin{equation}
\label{e15}
\begin{split}
 |\overrightarrow{p}|=&\, \sqrt{|\overrightarrow{p}|^2\left(1-f\left(\frac{|\overrightarrow{p}|}{E}\right)\right)+m^2} \simeq\\\,
\simeq
&
\, \, \, E\left(1-\frac{1}{2}f\left(\frac{|\overrightarrow{p}|}{E}\right)\right)+\frac{m^2}{2E}
\end{split}
\end{equation}
This procedure allows to evaluate the phase $\phi$ of the plane wave of eq.(\ref{e14}) for a given mass eigenstate, using the natural measure units, for which $t=L$:
\begin{equation}
\label{e16}
\phi=Et-EL+\frac{f}{2}EL-\frac{m^2}{2E}L=
\left(f E -\frac{m^2}{E} \right) \frac{L}{2}\, .
\end{equation}
Hence the same energy $E$ two mass neutrino eigenstates phase difference can be written as:
\begin{equation}
\label{e17}
\begin{split}
\Delta\phi_{kj} = & \, \phi_{j}-\phi_{k} =
\frac{(f_{j}-f_{k})}{2}EL-\left(\frac{m_{j}^{2}}{2E}-\frac{m_{k}^{2}}{2E}\right)L=\\
=&\left(\frac{\Delta m_{kj}^{2}}{2E}-\frac{\delta f_{kj}}{2}E\right)L
\end{split}
\end{equation}
In addition to the usual $3\times3$ unitary matrix PMNS, the oscillation probability shows therefore a dependence on the phase differences $\Delta\phi_{kj}$. In the most general case the transition probability from a flavor $|\alpha\rangle$ to a flavor $|\beta\rangle$,, that includes even the CP violating phase, can be written in the usual form:
\begin{equation}
\label{e18}
\begin{split}
P(\nu_{\alpha}\rightarrow\nu_{\beta}) = &\delta_{\alpha\beta}-4\sum_{i>j}\mathfrak{Re}\left(U_{\alpha i}U_{\beta i}^{*}U_{\alpha j}^{*}U_{\beta j}\sin^2(\Delta\phi_{ij})\right)+\\
+&2\sum_{i>j}\mathfrak{Im}\left(U_{\alpha i}U_{\beta i}^{*}U_{\alpha j}^{*}U_{\beta j}\sin^2(\Delta\phi_{ij})\right)
\end{split}
\end{equation}
The modified oscillation probability results modified and this effect is caused by the LIV violating perturbation term, proportional to $\delta f_{kj} = f_k - f_j$ in the phase differences defined in eq.(\ref{e17}). This term is different from zero only if the LIV violations coefficients $f_i$ are different for the three mass eigenstates. Otherwise the expression of equation (\ref{e18}) reduces to the usual three flavor oscillation probability, as in the case of absence of LIV.\\
It is essential to notice that in this LIV theory, MDR induced and CPT even, oscillation effects result caused by the difference of perturbations between different mass eigenstates (\cite{Liberati2}). The fundamental assumption, that represents a reasonable physical hypothesis, is that every mass state presents a personal maximum attainable velocity, since it interacts in a peculiar personal way with the background. It is even important to underline that the form of LIV, introduced in HMSR model, could not explain the neutrino oscillation, without resorting to the introduction of masses. In fact, the perturbative LIV mass term is proportional to the energy of the particle, and this is in contrast with the evidences of neutrino oscillations for the general pattern. In fact neutrino oscillations are well described by phase, depending only on squared masses differences, divided by the energy:
\begin{equation}
\label{e19}
\Delta\phi_{jk}=\left(\frac{m_{j}^2}{2E}-\frac{m_{k}^2}{2E}\right)=\frac{\Delta m_{jk}^2}{2E}L
\end{equation}
and LIV effects, of the type here introduced, could only appear at high energies as tiny perturbations (\ref{e17}). Therefore this model can account only for relatively  little deviations from ``standard physics'' and, at the highest observable energies in neutrino oscillation sector could generate only tiny perturbative effects. Nevertheless these effects are very interesting experimentally, because they could open a window on what can be new fundamental physics, the realm of quantum gravity.\\
Other LIV theories can explain oscillations, without resorting to the classical concept of neutrino masses \cite{Arias}.  They usually introduce terms in the Standard Model Lagrangian that generate masses by the interaction with background fields, as in \cite{Kostel3}, where the modified Dirac equation can be written using the modified Dirac matrices:
\begin{equation}
\label{e20}
\begin{split}
\Gamma_{AB}^{\,\mu}=&\gamma^{\mu}\delta_{AB}+c_{AB}^{\mu\nu}\gamma_{\nu}+d_{AB}^{\mu\nu}\gamma_{5}\gamma_{\nu}+\\
+&e^{\,\mu}_{AB}+if^{\mu}_{AB}\gamma_{5}+\frac{1}{2}g_{AB}^{\mu\nu\tau}\sigma_{\nu\tau}
\end{split}
\end{equation}
and the modified mass matrix:
\begin{equation}
\begin{split}
\label{e21}
M_{AB}=&m_{AB}+im_{5AB}\gamma_{5}+a^{\,\mu}_{AB}\gamma_{\mu}+\\
+&b^{\,\mu}_{AB}\gamma_{5}\gamma_{\mu}+\frac{1}{2}H_{AB}^{\mu\nu}\sigma_{\mu\nu}
\end{split}
\end{equation}
In the previous equations $m$ and $m_{5}$ are CPT and Lorentz symmetry preserving mass terms. The CPT preserving, but Lorentz violating terms are: $c,\,d,\,H$, while $a,\,b,\,e,\,f,\,g$ are CPT and consequently LI violating.
It is important to underline that in this case, the LIV introduced mass terms would constitute a theoretical justification for the oscillations, but this kind of LIV introduced masses would not modify the general dependence of oscillation probabilities on neutrino energy. Therefore, it would not amend the ``standard'' oscillation shape
with the introduction of new effects.
In order to evaluate the impact on neutrino phenomenology of the possible LIV, the three oscillation probabilities, ruling the neutrino oscillations ($P_{\nu_e\nu_{\mu}}$, $P_{\nu_e\nu_{\tau}}$  and $P_{\nu_{\mu}\nu_{\tau}}$) are evaluated by means of equations (\ref{e17}) and (\ref{e18})) in presence of LIV. Comparing the results with
the standard oscillation probabilities one gets if Lorentz invariance is satisfied.\\
This analysis has been pursued in the realistic three flavor scenario and the values of the $\Delta m^2_{ij}$ and of the various PMNS matrix elements ($U_{\alpha,i}$), used for the computations, have been taken from the most recent global fits, including all the different neutrino experiments \cite{Lisi-et-al,fit-altri}. For simplicity, the value $\delta = 0$ is assumed for the Dirac CP violation phase, this effect could be reintroduced, modifying in a simple way the analysis.\\
The outcome of the study on oscillation phenomenology is reported in the following series of figures. The different oscillation probabilities $P_{\nu_{\alpha}\nu_{\beta}}$ are plotted in absence and in presence of LIV violating terms. The plots are obtained for fixed neutrino beam energy values as a function of the baseline length L. The first series of $3$ graphs  are obtained for $E=1\,GeV$ and reports the probabilities $P_{\nu_{\mu}\nu_{\tau}},\,P_{\nu_{\mu}\nu_{e}}$ and $P_{\nu_{e}\nu_{\tau}}$.
The probability $P_{\nu_{\mu}\nu_{\tau}}$ is the most relevant one for the atmospheric neutrinos study and for long-baseline accelerator neutrino experiments. Even $P_{\nu_{\mu}\nu_{e}}$ is of great interest both for short and long baseline accelerator experiments and it is also important for reactor antineutrino experiments, because $P_{\bar{\nu}_{\mu}\bar{\nu}_{e}} = P_{\nu_{\mu}\nu_{e}}$ under the CPT invariance assumption.

\begin{figure}[H]
\includegraphics[width=135mm]{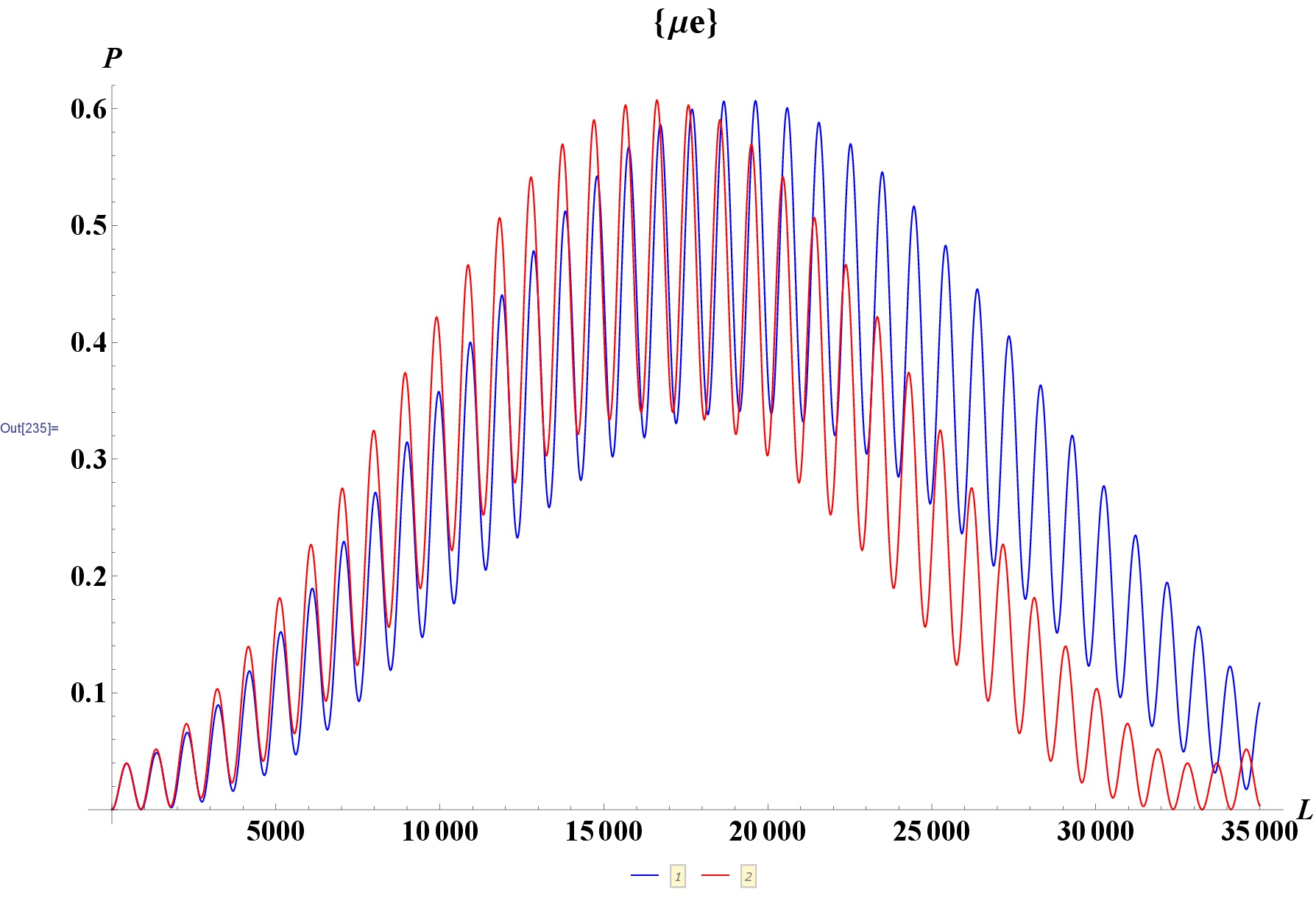}
\caption{Oscillation probability $\nu_{\mu}\rightarrow\nu_e$, computed for neutrino energy
${\rm E=1 \, GeV}$, ``standard theory" (red curve) and LIV (blue curve), for LIV parameters $\delta f_{32} = \delta f_{21} = 1 \times 10^{-23}$, as function of the baseline $L$. \cite{Antonelli}}
\label{mue1gev}
\end{figure}
\begin{figure}[H]
\includegraphics[width=135mm]{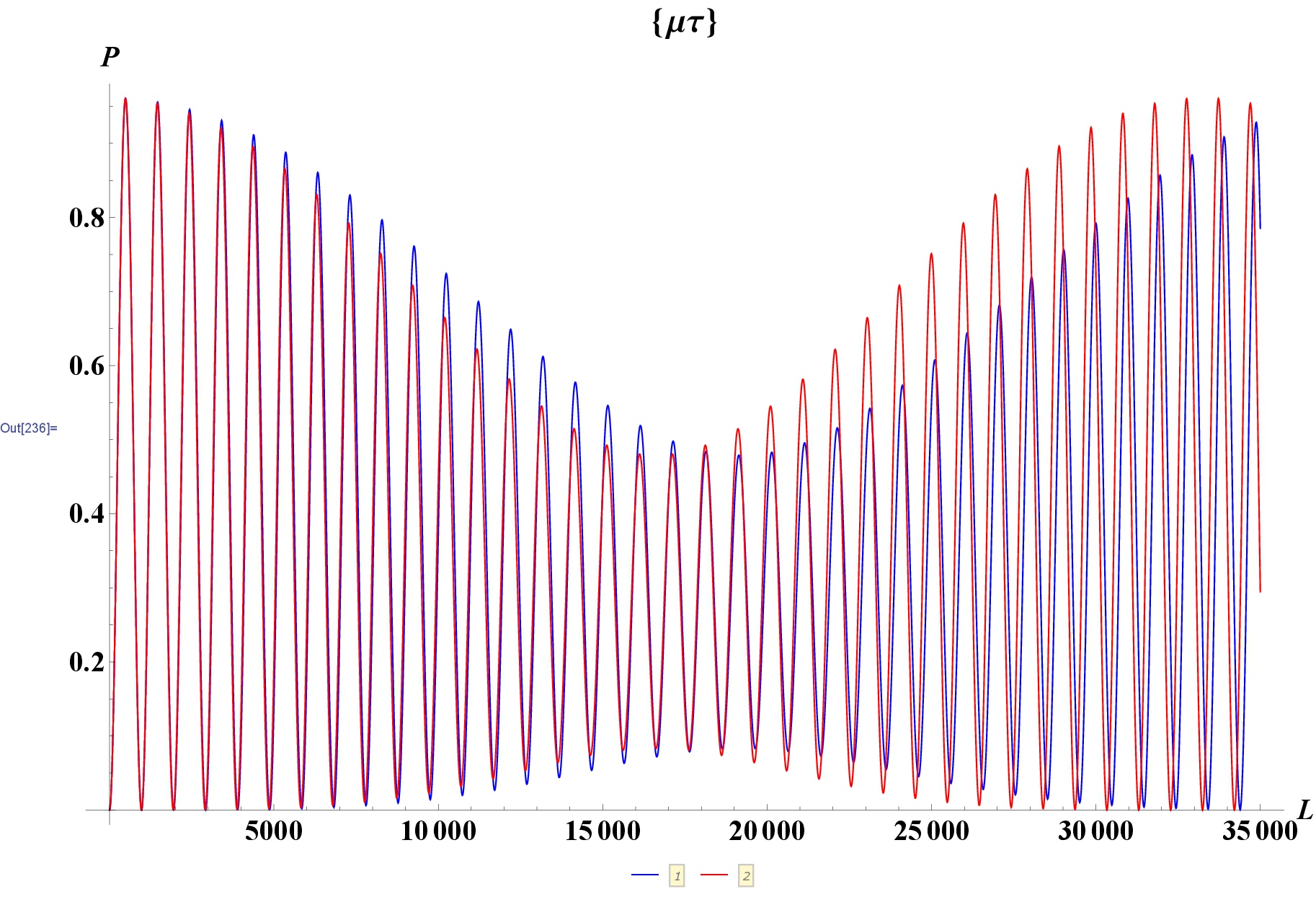}
\caption{Same analysis of  fig.\ref{mue1gev}, but for the oscillation $\nu_{\mu}\rightarrow\nu_{\tau}$. \cite{Antonelli}}
\label{mutau1gev}
\end{figure}
\begin{figure}[H]
\includegraphics[width=135mm]{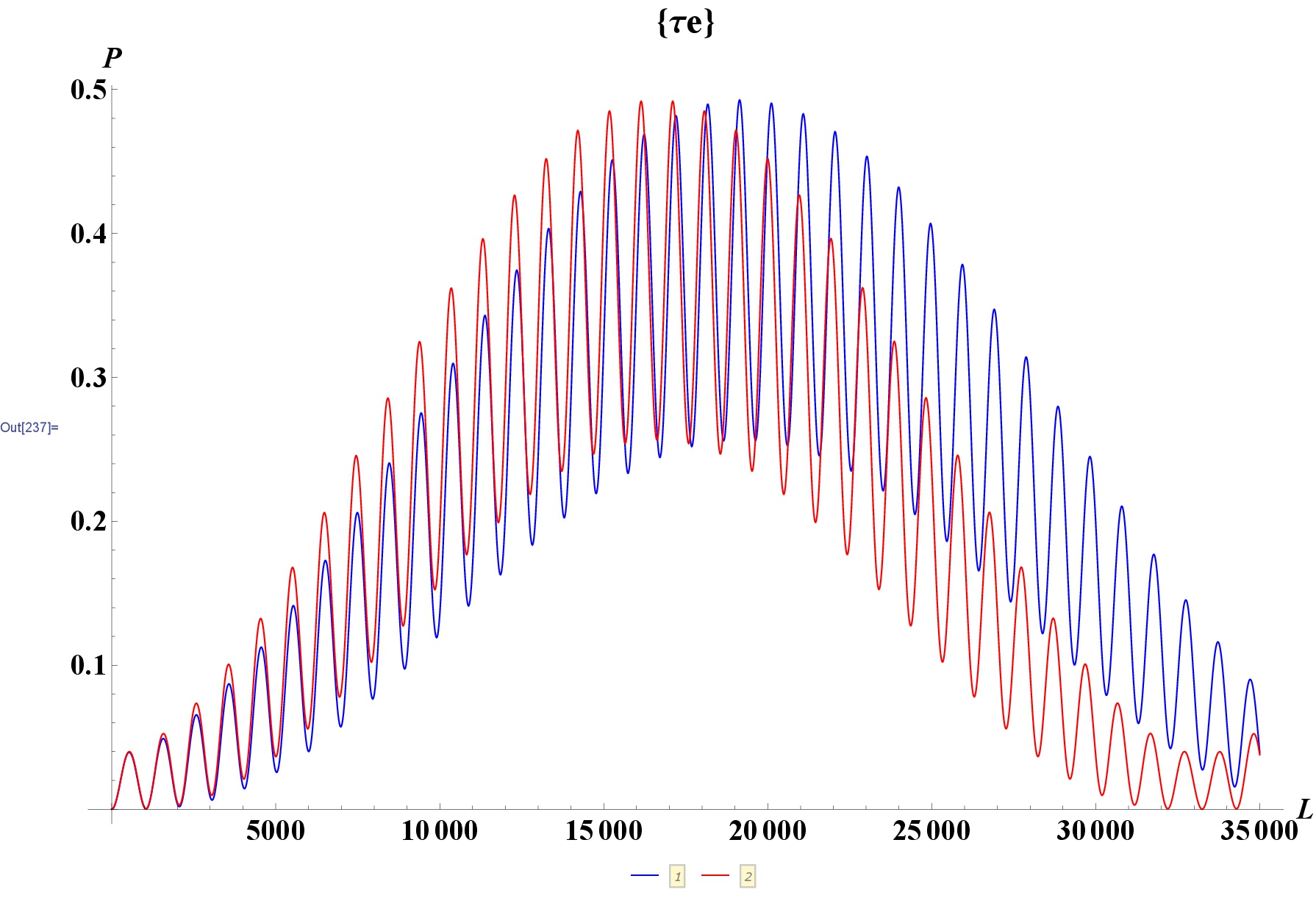}
\caption{Same analysis of  fig.\ref{mue1gev}, but for the oscillation $\nu_{e}\rightarrow\nu_{\tau}$. \cite{Antonelli}}
\label{etau1gev}
\end{figure}

The $3$ LIV correction parameters $f_{k}$ are assumed of the same magnitude and are ordered with the highest associated to the heaviest mass eigenstate, following a natural order. The perturbation magnitude is governed by the differences $\delta f_{32}$ and $\delta f_{21}$ (\ref{e16}) In figs.\ref{mue1gev}-\ref{etau1gev} the values $\delta f_{32} = \delta f_{21}= 1 \times 10^{-23}$ are employed and for energy beam of $E=1\,GeV$ LIV would modify in a visible way the oscillation probabilities patterns.\\
The parameters introduced in this work present some differences from other analysis. HMSR in fact investigates a sector not yet considered in other models, so the comparison between this model and other results present in literature is not so immediate \cite{Russell} (result obtained for instance by SuperKamiokande collaboration employing SME Hamiltonian approach \cite{SK-test-LIV}).

\begin{figure}[H]
\includegraphics[width=135mm]{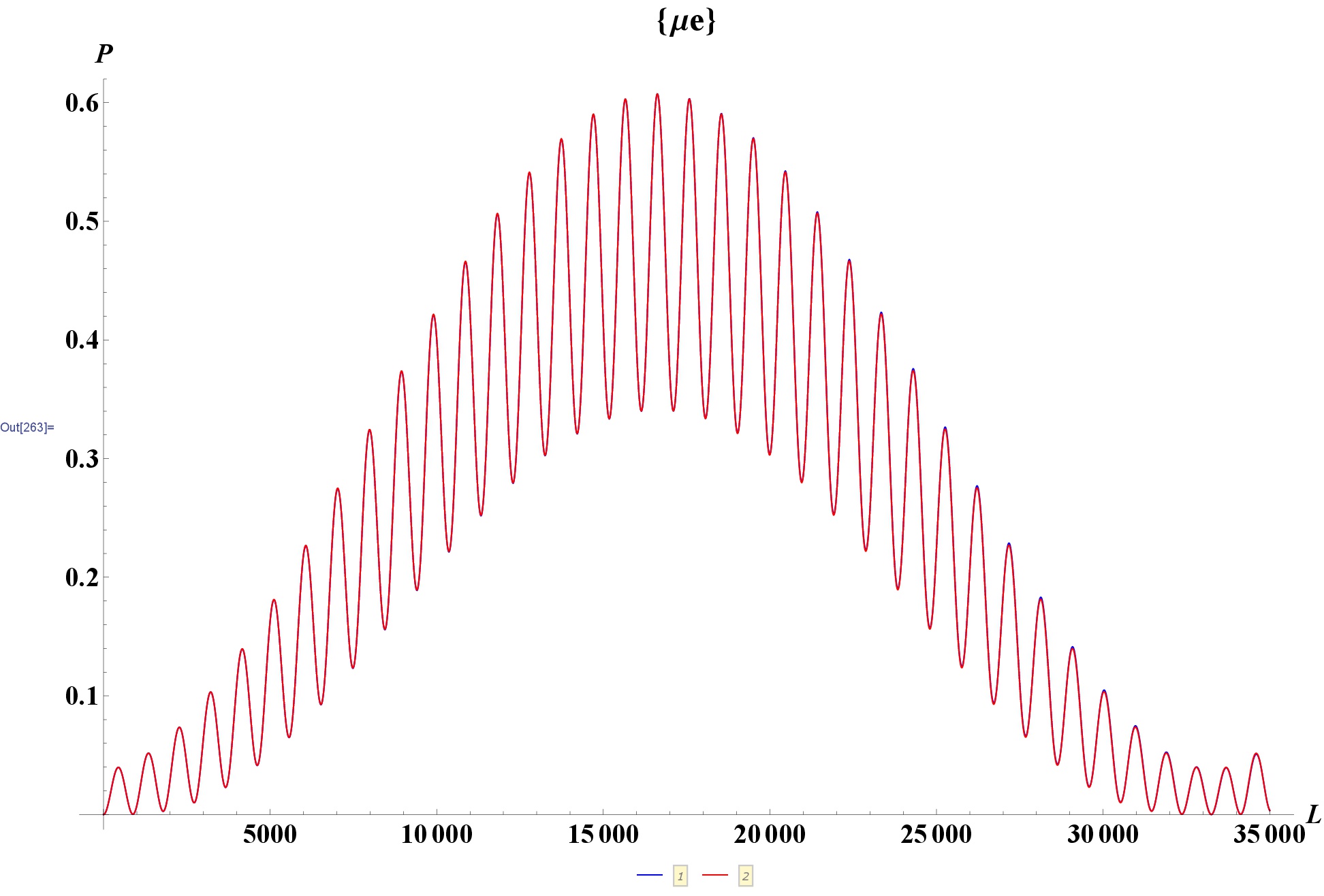}
\caption{Example of the fact that with LIV of parameters $\delta f_{kj}\simeq10^{-25}$ for energy beam of $E=1\,GeV$ LIV effects are not visible. \cite{Antonelli}}
\label{mue1gevpiccolo}
\end{figure}

\begin{figure}[H]
\includegraphics[height=90mm]{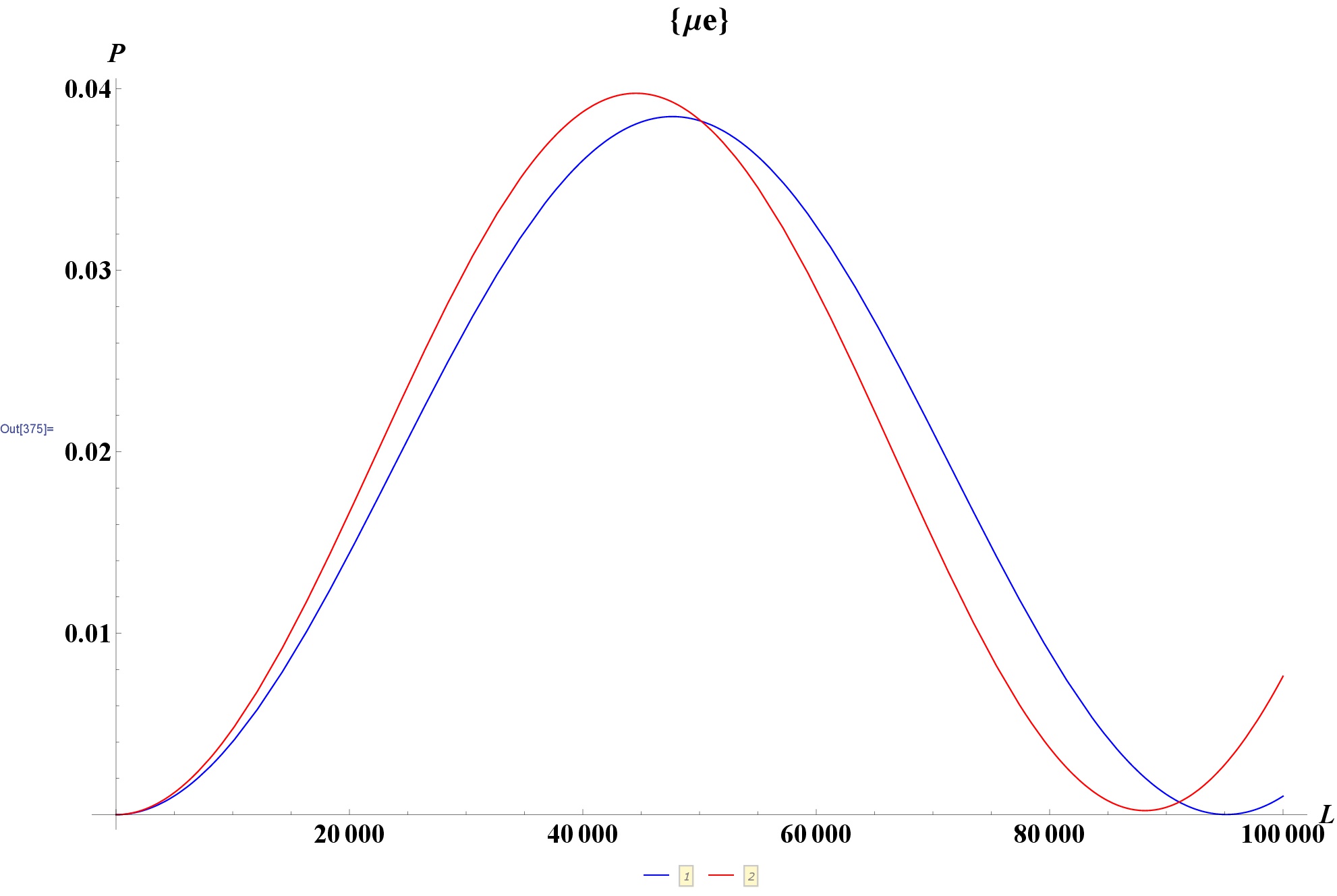}
\caption{Same analysis of fig.\ref{mue1gev}, but for LIV parameters $\delta f_{32} = \delta f_{21} = 4.5 \times 10^{-27}$   and for neutrino energy beam {\rm E = 100 GeV}. \cite{Antonelli}}
\label{mue100gev}
\end{figure}

\begin{figure}[H]
\includegraphics[height=90mm]{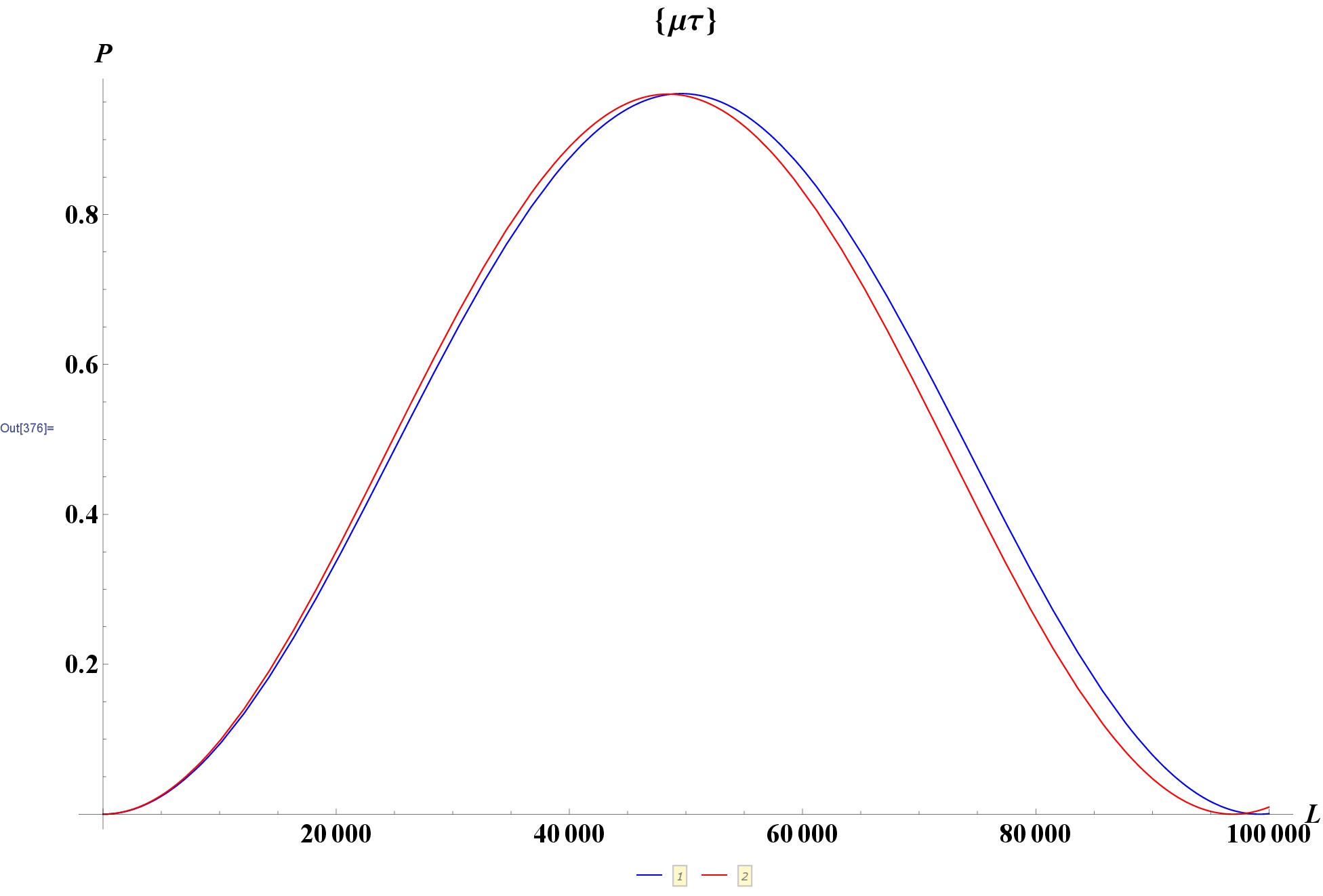}
\caption{Same of  fig.\ref{mue100gev} in the case of the oscillation probability
$P_{\nu_{\mu}\nu_{\tau}}$. \cite{Antonelli}}
\label{mutau100gev}
\end{figure}

\begin{figure}[H]
\includegraphics[height=90mm]{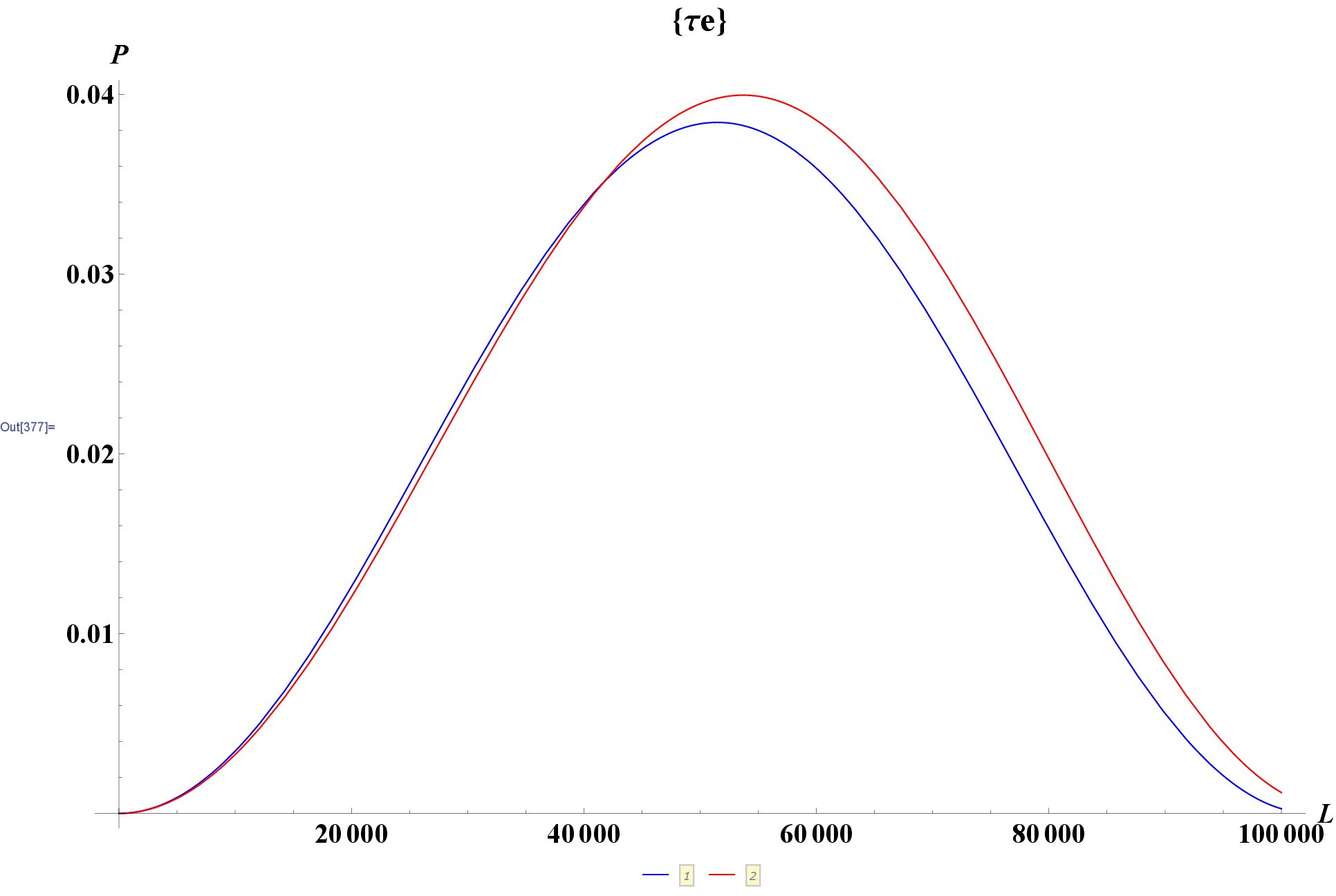}
\caption{Same of  fig.\ref{mue100gev}, but for $P_{\nu_{e} \nu_{\tau}}$. \cite{Antonelli}}
\label{etau100gev}
\end{figure}

For lower LIV parameters magnitude effects are visible only for higher energy beam values. In figs. \ref{mue100gev}-\ref{etau100gev} the results for the $3$ oscillation probabilities are plotted, in the case of $E=100\,GeV$ energy beams. In these plots the LIV parameters are assumed in order to obtain $\delta f_{32}=\delta f_{21}=4.5\times10^{-27}$ and perturbation effects are visible. Even the effects for $E=1\,TeV$ neutrino are studied. Neutrino energies in the region from $TeV$ to $PeV$ are of great interest for neutrino telescopes experiments like ANTARES \cite{ANTARES}, KM3NET \cite{KM3NET}, IceCube \cite{IceCube} and Auger \cite{Auger} (the last one for cosmic neutrinos with energies above $EeV$). In figs. \ref{Pmutau1tev3par}-\ref{Petau1tev3par} are reported the results for $E=1\,TeV$ energy neutrinos, obtained for various LIV magnitude parameters.

\begin{figure}[H]
\includegraphics[width=135mm]{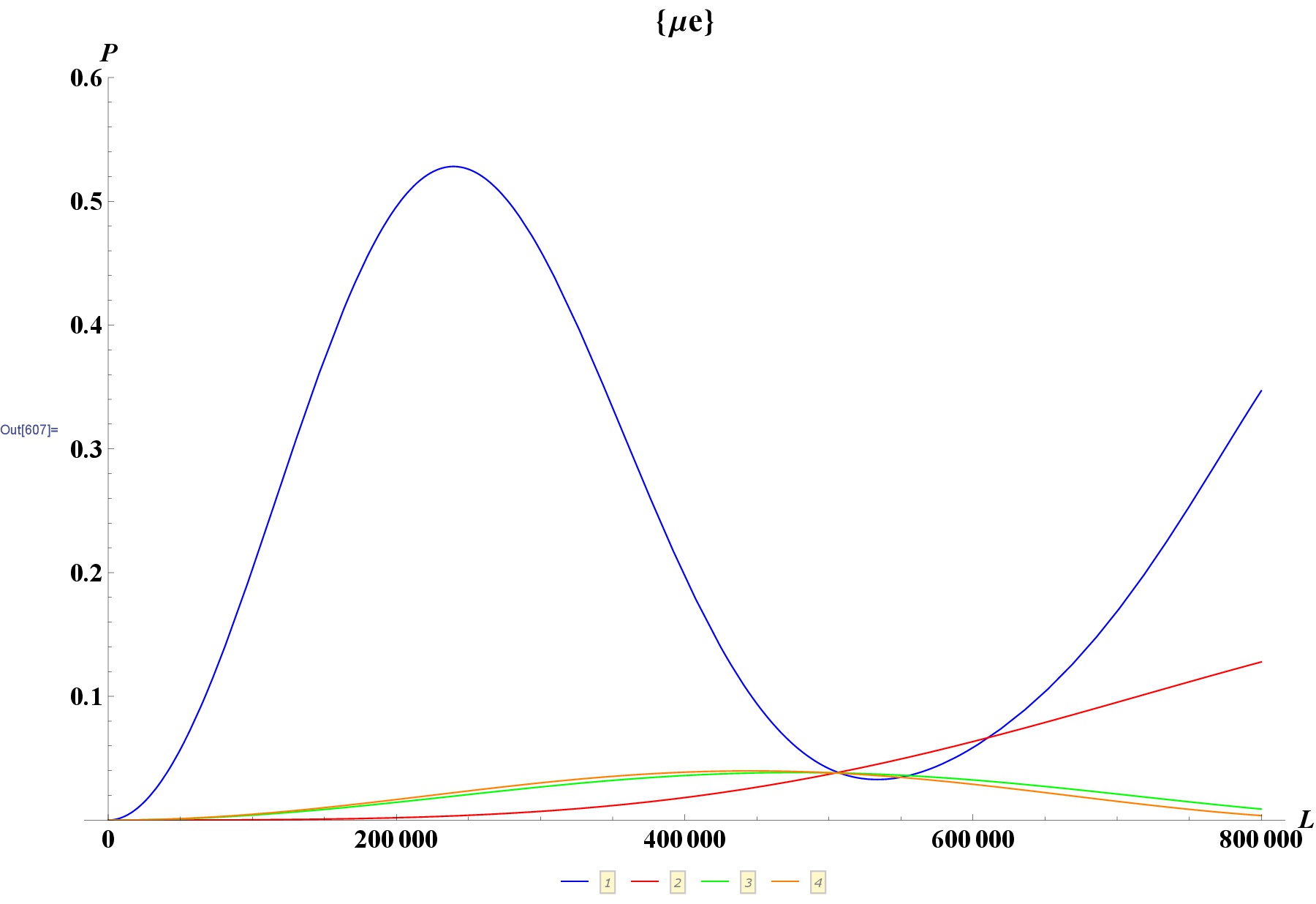}
\caption{$P_{\nu_{\mu}\nu_e}$ oscillation probability, as function of baseline L, for neutrino energy ${\rm E = 1 \, TeV}$, for
"classical theory", LI (orange curve) and for LIV models, with parameters equal respectively
to $\delta f_{32} = \delta f_{21} = 4.5 \times 10^{-27}$ (blue), $\delta f_{32} = \delta f_{21} = 4.5 \times 10^{-28}$ (red) and $\delta f_{32} = \delta f_{21} = 4.5 \times 10^{-29}$ (green curve). \cite{Antonelli}}
\label{Pmue1tev3par}
\end{figure}

\begin{figure}[H]
\includegraphics[width=135mm]{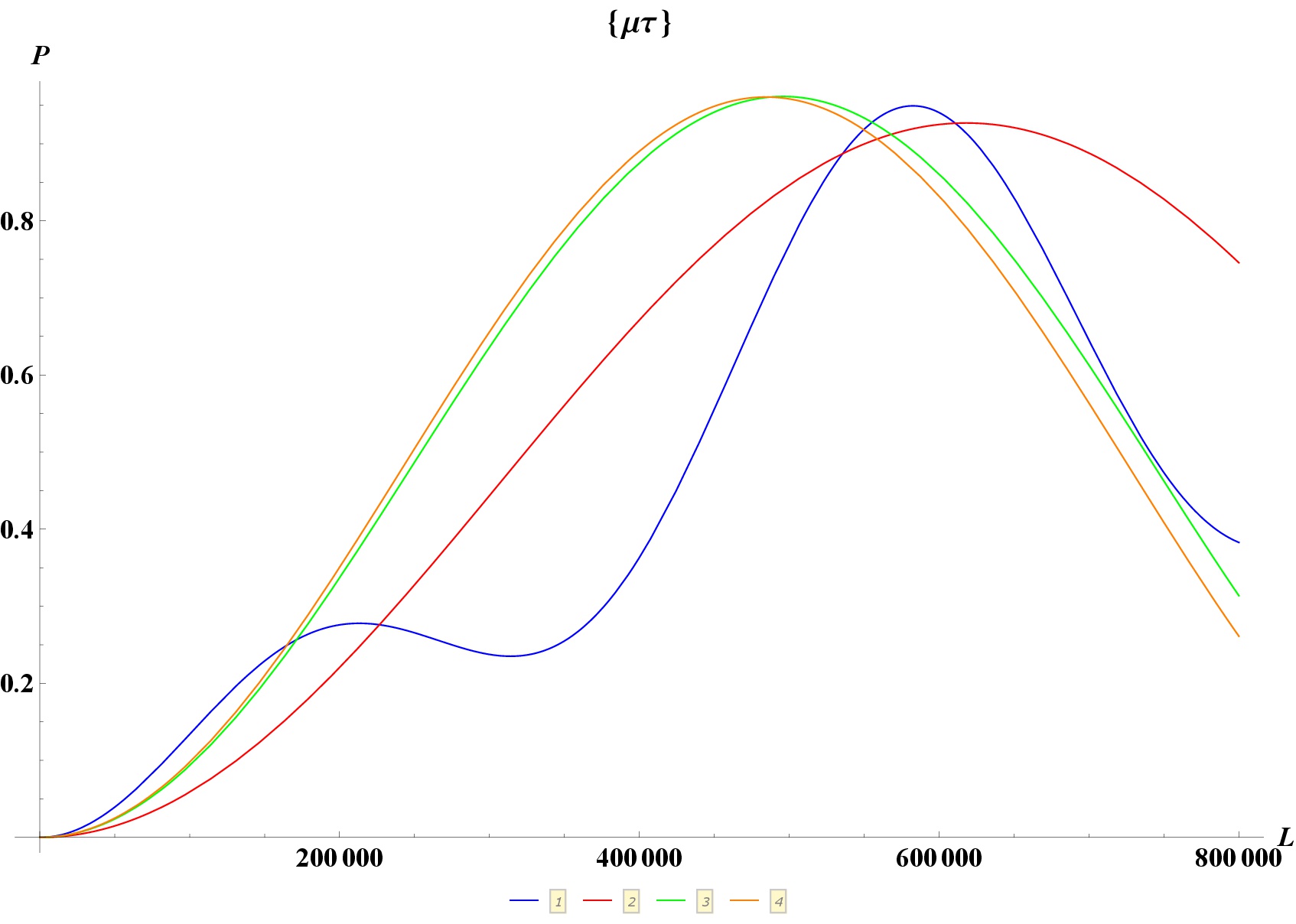}
\caption{Same analysis of fig.\ref{Pmue1tev3par}, but for the case of $P_{\nu_{\mu}\nu_{\tau}}$. \cite{Antonelli}}
\label{Pmutau1tev3par}
\end{figure}

\begin{figure}[H]
\includegraphics[width=135mm]{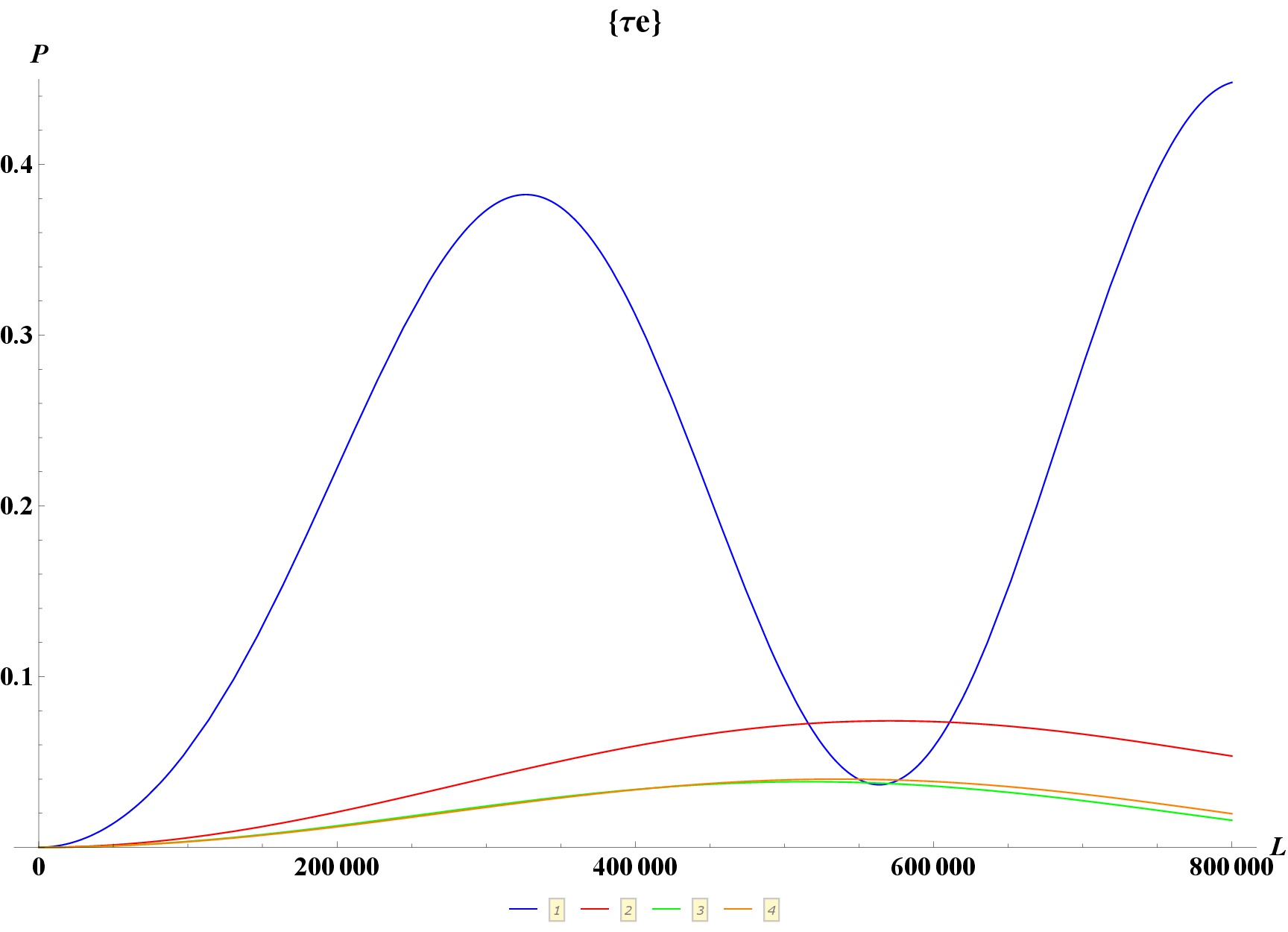}
\caption{Same analysis of fig.\ref{Pmue1tev3par}, but for $P_{\nu_{e}\nu_{\tau}}$. \cite{Antonelli}}
\label{Petau1tev3par}
\end{figure}

\begin{figure}[H]
\includegraphics[width=150mm]{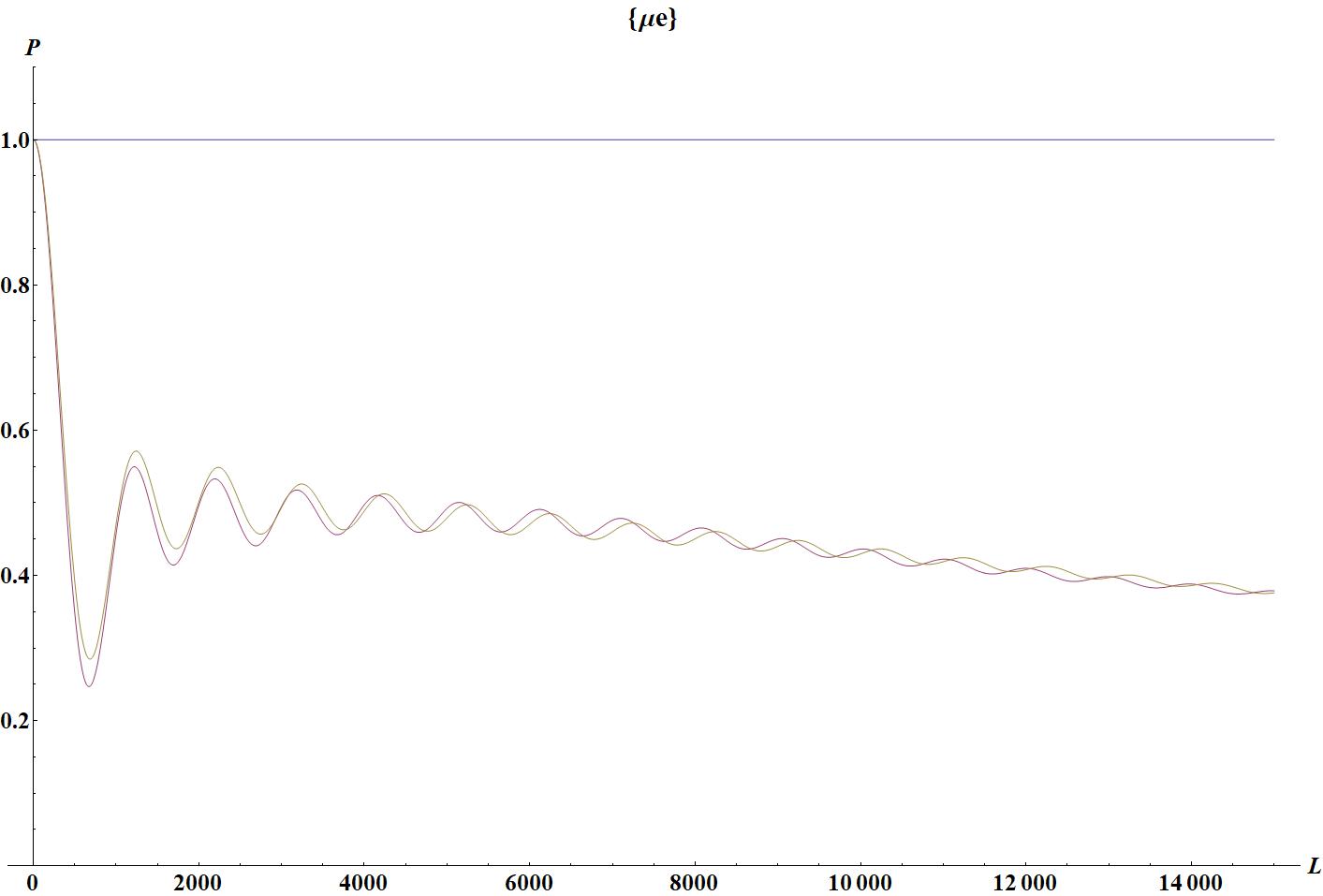}
\caption{Comparison between oscillation probability integral averaged on energy range values from $1\,GeV$ to $10\,GeV$, in LI scenario (blue line) and LIV scenario (red line), with $\delta f_{32} =10^{-23}$ and $\delta f_{21} =10^{-25}$, baseline $L$ in km.}
\label{110}
\end{figure}

\begin{figure}[H]
\includegraphics[width=150mm]{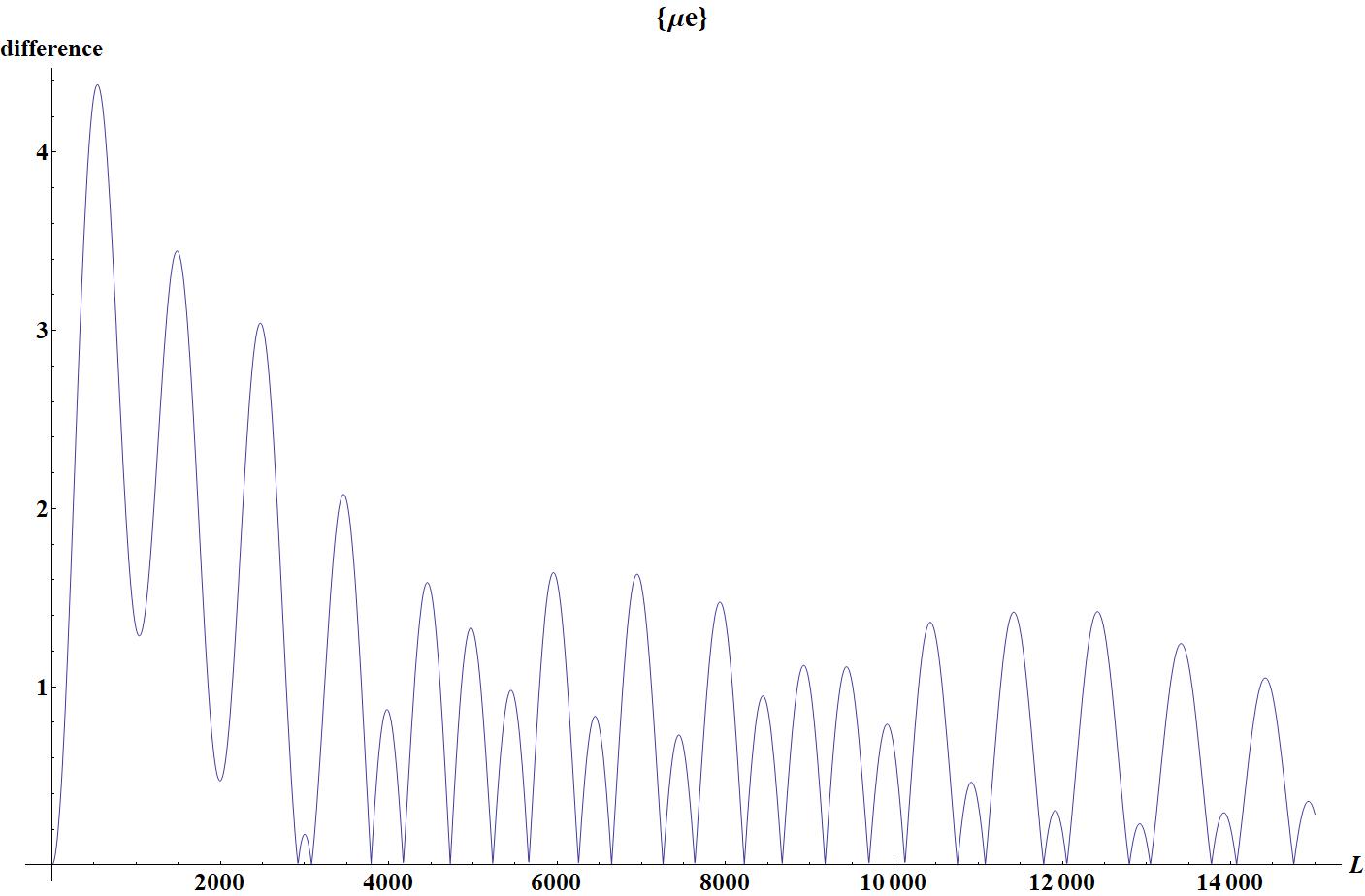}
\caption{Percentage difference of the two probabilities from the previous plot (LI and LIV), baseline $L$ in km.}
\label{111}
\end{figure}

Hence, selecting the appropriate experimental context, in future one could use the detailed study of high energy neutrinos to further constraint the LIV coefficients.
To obtain a realistic phenomenological analysis, useful for realistic experimental scenarios, one needs to take into account the knowledge of the different interaction energy depending cross sections $\sigma_{\beta} (E)$ of a $\beta$ neutrino with the detector and an accurate knowledge of $\Phi_{\alpha} (L, E)$ the foreseen initial flux of an $\alpha$ flavor neutrino at given energy $E$, to integrate the information derived from probability.
The number $N_{\alpha,\beta}$ of detected transition events caused by the $\nu_{\alpha} \to \nu_{\beta}$ flavor oscillation, will be given by:
\begin{equation}
\label{e22}
N_{\alpha,\beta} \propto \Phi_{\alpha} (L, E)  \, P_{\nu_{\alpha},\nu_{\beta}} (L, E) \, \sigma_{\beta} (E)
\end{equation}
where $L$ represents the distance from the production to the detection point. Then this information must be integrated over the neutrino energies and finally one must take into account functions describing the detector resolution and efficiencies.\\
It must be underlined that the detector can not have a point resolution in energy, so it is important to report a plot indicating a comparison between the oscillation probability integral averaged respect to a range of energy values in fig. \ref{110}. To complete the analysis even the percentage differences of LI and LIV predicted integral averaged probabilities are is plotted in fig.\ref{111}. In these plots the maximum baseline $L$ has been chosen in order to be of interest for atmospheric neutrino analysis.

The effects of LIV are visible and in the case of higher energies involved or increasing sensitivity of the detector it is possible to pose more restrictive constrains on the LIV parameters.\\
From the comparison between the experimental results and the theoretical predictions, one can extract the information about the impact of this model supposed LIV violations. Otherwise one can put constraints on the magnitude order of the LIV coefficients.

\section{LIV and Mass Hierarchy}
Another interesting work about LIV and neutrino oscillation is that of Jurkovich \cite{Jurkovich}. In that work the model investigated is based on the SME, so it does not preserve the covariance of the theory respect to amended Lorentz transformations. Moreover the effects of LIV not only on oscillations, but even on the possibility to detect the neutrino Mass Hierarchy (MH) are investigated and what emerges is that LIV can affect the long base experiments sensitivity to this kind of detection.\\
The neutrino sector is investigated introducing a modified Lagrangian that introduces changes in the kinematical terms:
\begin{equation}
\label{z1}
L_{d-dim}=i\nu^{\dagger}_{iL}\partial_{\mu}\nu_{iL}-i^{d-3}\gamma_{i}^{j_{1}...j_{d-4}}\nu_{iL}^{\dagger}\sigma^{k}\partial_{k}\partial_{j_{1}}...\partial_{j_{d-4}}\nu_{iL}
\end{equation}
where $\gamma_{i}^{j_{1}...j_{d-4}}$ are $d-4$ tensors and $\sigma^{k}$ are the Pauli matrices.
The dispersion relations are modified and again assume the form:
\begin{equation}
\label{z2}
E^2=(1+\overline{\gamma})^2 \textbf{p}^2
\end{equation}
where $\overline{\gamma}=\gamma_{i}^{\,j_{1}...j_{d-4}}p_{j_{1}}...p_{j_{d-4}}$
If massive neutrinos are considered the dispersion relation assumes the explicit form:
\begin{equation}
\label{z3}
E^2=(1+\overline{\gamma})^2 \textbf{p}^2+m^2
\end{equation}
and the usual Hamiltonian in the mass basis assumes the explicit form:
\begin{equation}
\label{z4}
H\longrightarrow H_{0}+H_{LIV}
\end{equation}
where $H_{0}$ represents the usual Hamiltonian:
\begin{equation}
\label{z5}
H_{0}=\left(
        \begin{array}{ccc}
          0 & 0 & 0 \\
          0 & \Delta m_{12}^2/2E & 0 \\
          0 & 0 & \Delta m_{31}^2/2E \\
        \end{array}
      \right)+ U\,V(x)\,U^{\dagger}
\end{equation}
$H_{LIV}$ represents the perturbation term introduced by LIV:
\begin{equation}
\label{z6}
H_{LIV}=\left(
          \begin{array}{ccc}
            0 & 0 & 0 \\
            0 & \Delta\gamma_{21}^{d}E^{d-3} & 0 \\
            0 & 0 & \Delta\gamma_{31}^{d}E^{d-3} \\
          \end{array}
        \right)
\end{equation}
Now the results of \cite{Jurkovich} are compared with those produced by HMSR applied to the MH detection for experiments like JUNO \cite{An}. Resorting to the model presented in the previous section and used for investigating the effects of LIV on oscillation, the spectrum of the foreseen detected neutrinos is constructed. What is possible to see is that LIV can not affect the general shape of the foreseen spectrum and the shape of direct and inverse hierarchy oscillations. Therefore it is possible to conclude that short and medium baseline experiments as JUNO can detect LIV effects in an efficient way only studying the neutrino probability and not mass hierarchy. This result was expected, since reactor neutrino have energy with magnitude order of $MeV$ and LIV perturbations start to be visible for $GeV$ neutrino beams. In fact the phase perturbation introduced by HMSR is proportional to $L\times E$.

\begin{figure}[H]
\includegraphics[width=155mm]{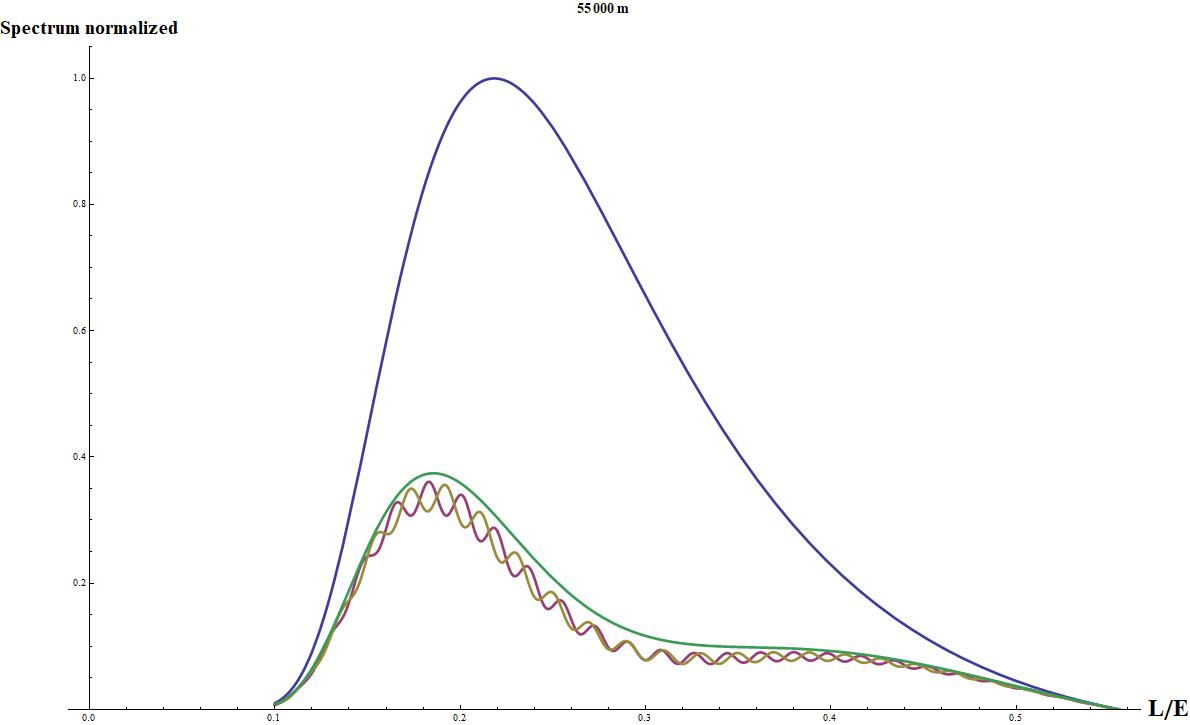}
\caption{Foreseen spectrum for JUNO experiment plotted as a function of the variable L/E, for fixed baseline L\,=\,55\,km.}
\label{aaa}
\end{figure}

\begin{figure}[H]
\includegraphics[width=155mm]{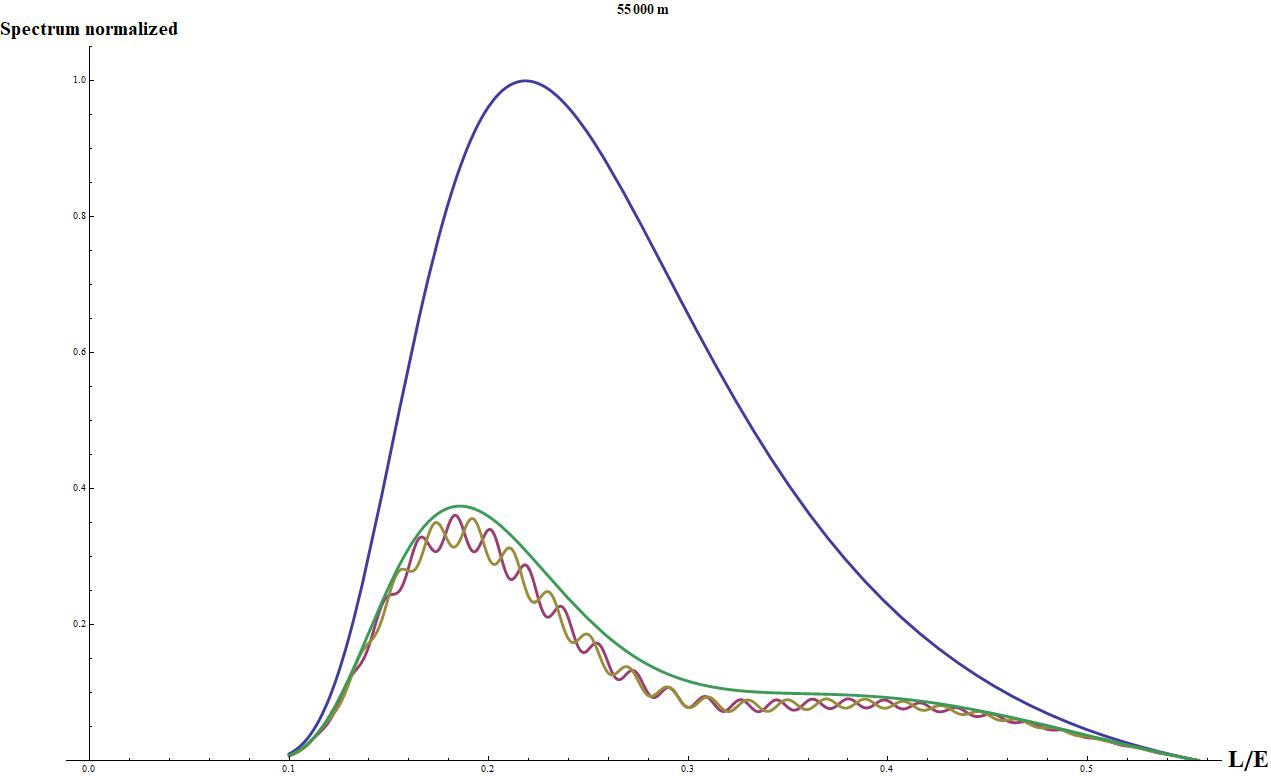}
\caption{Modified spectrum by LIV for JUNO experiment, , with $\delta f_{32} =10^{-23}$ and $\delta f_{21} =10^{-25}$, plotted as a function of the variable L/E, for fixed baseline L\,=\,55\,km. In this plot there are no visible differences respect to the previous one. }
\label{bbb}
\end{figure}

\section{Conclusions}
Neutrino physics is an ideal playground to search for deviations from Lorentz invariance, thanks to its various set of experiments, covering a wide spectrum of energies and baselines. Other works for instance explore the influence of LIV on the foreseen of observed spectra, in the case of superluminal correction to neutrino propagation \cite{Stecker,Carmona}. Short and Long baseline neutrino experiments seem to be the ideal structures to test the validity of Lorentz Invariance, due to their great sensitivity to the detection of phase differences in neutrino propagation. So it is interesting to consider future experiments such as JUNO, DUNE, T2K to constrain the magnitude of LIV perturbations. In particular it will be interesting to conduct a systematic analysis of what can be detected by new and really advanced facilities, like JUNO for instance, regarding the study of neutrino oscillations probability. Even the investigation on the effects of LIV on the Mass Hierarchy discrimination for long baseline experiments can be an interesting research aim. In fact it can open another window on the study of fundamental symmetries of nature, presenting another sector where posing Lorentz Invariance under investigation.

%%%%%%%%%%%%%%%%%%%%%%%%%%%%%%%%%%%%%%%%%%

%%%%%%%%%%%%%%%%%%%%%%%%%%%%%%%%%%%%%%%%%%
\vspace{15pt}

%%%%%%%%%%%%%%%%%%%%%%%%%%%%%%%%%%%%%%%%%%
%% optional
%\supplementary{The following are available online at \linksupplementary{s1}, Figure S1: title, Table S1: title, Video S1: title.}


%%%%%%%%%%%%%%%%%%%%%%%%%%%%%%%%%%%%%%%%%%
\textbf{Funding:} 
This work was supported by the Fondazione Fratelli Confalonieri - Milano
%%%%%%%%%%%%%%%%%%%%%%%%%%%%%%%%%%%%%%%%%%
\textbf{Acknowledgments:}
The author would like to thank Lino Miramonti and Vito Antonelli for the great opportunity of working together and learning from them useful concepts about neutrino physics.

\textbf{Conflicts of interest:}
The author declares no conflict of interest.

\end{document}